\documentclass[pdflatex,sn-mathphys]{sn-jnl}

\usepackage{soul}
\usepackage{amsmath}
\usepackage{amssymb}
\usepackage{amsfonts}
\usepackage{url}

\jyear{2022}%

\theoremstyle{thmstyleone}%
\theoremstyle{thmstyletwo}%

\theoremstyle{thmstylethree}%

\begin{document}

\title[Data-driven Parametric Insurance Framework]{Data-driven Parametric Insurance Framework Using Bayesian Neural Networks}


\author*[1,3]{\fnm{Subeen} \sur{Pang}}\email{sbpang@mit.edu}
\equalcont{These authors contributed equally to this work.}

\author[2,3]{\fnm{Chanyeol} \sur{Choi}}\email{chanyeol@mit.edu}
\equalcont{These authors contributed equally to this work.}

\affil[1]{\orgdiv{Department of Mechanical Engineering}, \orgname{Massachusetts Institute of Technology (MIT)}, \city{Cambridge}, \state{MA}, \country{United States}}

\affil[2]{\orgdiv{Department of Electrical Engineering and Computer Science}, \orgname{Massachusetts Institute of Technology (MIT)}, \city{Cambridge}, \state{MA}, \country{United States}}

\affil[3]{\orgname{Wecover Platforms, Inc., \city{Cambridge}, \state{MA}, \country{United States}}}


\abstract{As climate change poses new and more unpredictable challenges to society, insurance is an essential avenue to protect against loss caused by extreme events. Traditional insurance risk models employ statistical analyses that are inaccurate and are becoming increasingly flawed as climate change renders weather more erratic and extreme. Data-driven parametric insurance could provide necessary protection to supplement traditional insurance. We use a technique referred to as the deep sigma point process, which is one of the Bayesian neural network approaches, for the data analysis portion of parametric insurance using residential internet connectivity dropout in US as a case study. We show that our model has significantly improved accuracy compared to traditional statistical models. We further demonstrate that each state in US has a unique weather factor that primarily influences dropout rates and that by combining multiple weather factors we can build highly accurate risk models for parametric insurance. We expect that our method can be applied to many types of risk to build parametric insurance options, particularly as climate change makes risk modeling more challenging.}




\maketitle

\section{Introduction}\label{sec1}

Extreme weather is a major challenge that will continue to evolve and worsen as climate change progresses. Stronger and less predictable storms, worse droughts and floods, extreme temperatures, and rising sea levels affect societies around the globe \cite{chester2020keeping, neumann2015climate, yates2014stormy}. These effects reach many vital facets of modern society, from agriculture to transportation to infrastructure, to name a few \cite{falco2014crop, ji2022impact, jorgensen2020natural, national2008potential, wang2012impact, wilbanks2013climate}. As we work to stall and reverse climate change, insurance offers an avenue to protect those who are affected by the devastating effects of extreme weather \cite{authority2015impact, botzen2010climate, falco2014crop, mills2007synergisms}.

Studies assessing how insurance can best support individuals and societies during climate change highlight the importance of new models and products. Studies primarily focus on solutions that fall into three categories: building new models that are more accurate, working to help mitigate the direct effects of climate change, and advocating politically for policies that ameliorate climate change \cite{collier2021climate, dlugolecki2008climate, hecht2007climate, mills2009global}. While the second two methods are essential steps towards mitigating and reversing the impacts of climate change, the first offers an opportunity to use insurance itself to protect individuals and societies from the direct damages. Looking at how insurance companies traditionally predict and underwrite risk gives us an opportunity to assess how these models can be improved. 

Currently, most of the insurance products in the field of climate-based risk are indemnity-based \citep{kraehnert2021insurance}. In addition, multiple studies have demonstrated mathematical models to estimate climate-based risk, which is closely related to such products, ranging from the well-known linear regression \cite{pan2022assessing}, Bayesian modeling \cite{lawrence2020leveraging}, and machine learning techniques \cite{ma2019flash} to even the latest methods on neural networks \cite{cesarini2021potential}. However, these models are only available to limited cases. For example, it is common that the relationship between climate conditions and corresponding risk is highly nonlinear, making regression models likely to be inaccurate or incomplete. Machine learning techniques such as support vector machines and gradient boosting methods could results in robust outcomes, but they are only applicable to certain priors or data structures. Artificial intelligence (AI) based models, especially neural networks, are highly flexible and have shown remarkable results in many branches of fields. However, they may cause over-fitting issues and they do not provide the uncertainty of their estimations (like the posterior variance in Bayesian modeling). In addition to the type of models, data pre-processing becomes more significant. In particular, insurance risk models are focused on extreme cases, i.e. long-tail statistics, it becomes more difficult to accurately estimate risk models due to the sensitivity to noises and data sparsity. Therefore, feeding AI risk models with raw dataset without pre-processing could lead to non-comprehensive results.

On the other hand, parametric insurance has been proposed as a supplement to traditional insurance, because it can mitigate expensive claim procedures and so-called moral hazard issues \citep{porrini2014insurance,mahul2010government,vroege2019index,miranda2012index}. This can be critical in developing countries and rural areas where documentation of damages in properties and communication with policyholders are difficult and costly. However, the existence of basis risk -- mismatch between risk estimations and policyholders' actual losses -- remains as one of the main problems in parametric insurance. Hence, robust analysis on extreme events becomes a critical component in parametric insurance, minimizing basis risk.

For the minimization of basis risk and better estimation of consequences from extreme weathers, we propose using the deep sigma point process (DSPP) \cite{jankowiak2020deep}, which is a variant of the deep Gaussian process (DGP) \cite{damianou2013deep}. In AI studies, DGP is classified as a technique in Bayesian neural network (BNN) framework. BNNs are different from conventional neural networks, because they can estimate uncertainty regarding its estimations. For example, conventional networks can always tell us how risky a weather condition is, but such decision may not be reliable due to insufficient amount of examples or unstable training process. In contrast, BNNs can estimate model's confidence, and we expect that this can be useful in insurance business, since we may apply more conservative policies on weather conditions with high uncertainty. In addition, BNNs are more robust to overfitting. This is because the deterministic and fixed estimation of parameters in conventional neural networks is now substituted by probability distributions that represent uncertainty from lack of data, optimal model architecture, etc \cite{cutajar2018deep}. There are a few advantages that we specifically choose DSPP (or DGP) over other techniques in the BNN framework. One representative example is that Gaussian processes can represent uncertainty directly in function-space, which can lead to better estimations than other BNNs with weight-space probability distributions \cite{foong2020expressiveness,damianou2015deep}. 

\begin{figure}[t]%
    \centering
    \includegraphics[width=0.8\textwidth]{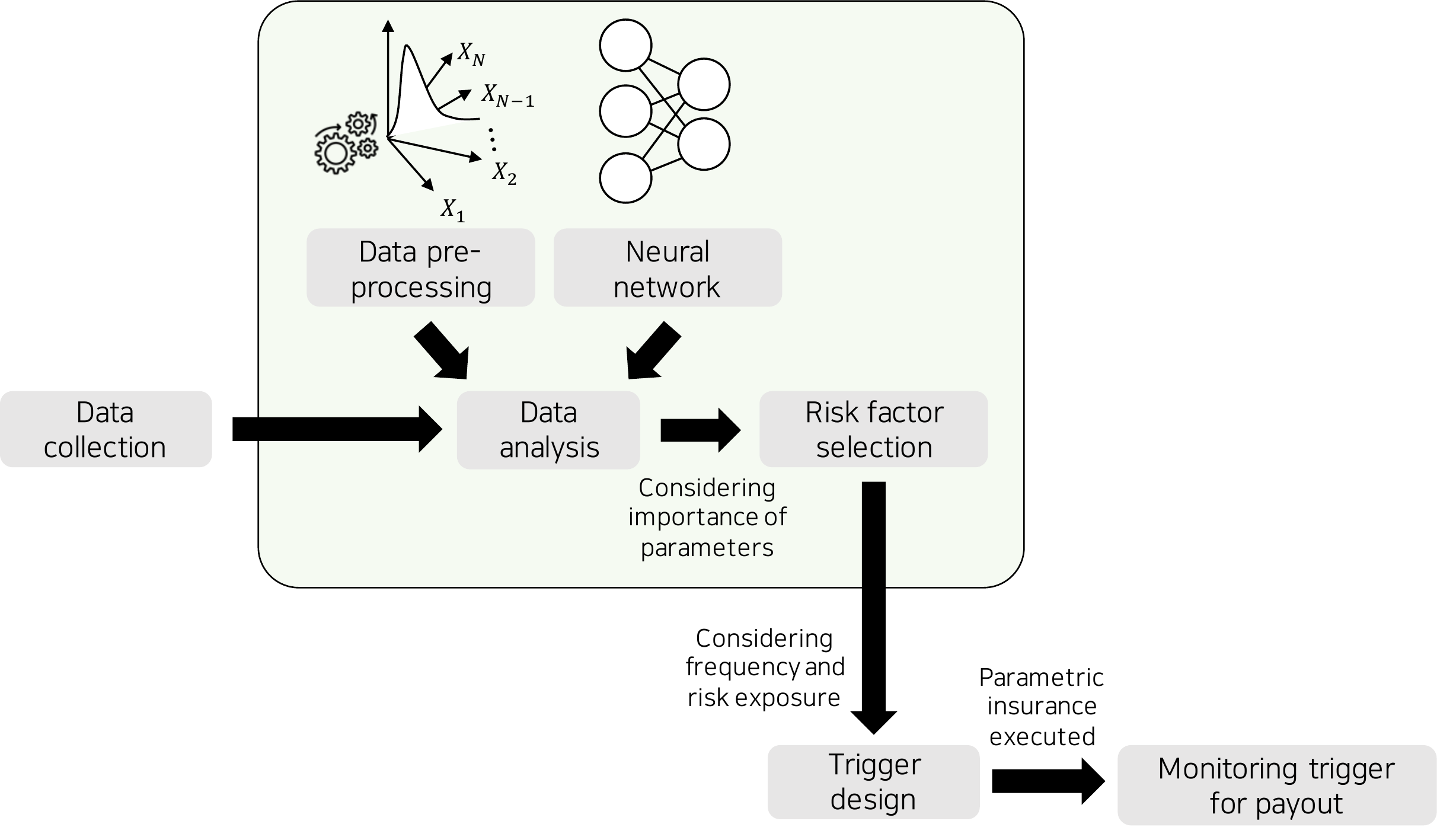}
    \label{fig:overview}
    \caption{Overview of the data-driven parametric insurance. The first step of parametric insurance is data collection, which is followed by data analysis. In the boxed region, we show that our method applies data pre-processing and neural networks to the data analysis stage. Risk factors are then selected for the trigger design. From this point the parametric insurance is executed and the triggers are monitored for payout.}
\end{figure}

Given the advantages of DSPPs while leveraging the high flexibility and expressiveness of neural networks, we hypothesize that we could use DSPPs for more accurate parametric insurance. We develop a framework including DSPPs to create a model that can predict the effect of weather parameters on dropout in residential links, using a public dataset collected by ThunderPing system \cite{schulman2011pingin} over 8 years across the US \cite{padmanabhan2019residential}. This data allow us to build and assess the ability of our models to reliably predict risk. In this study, we demonstrate that DSPPs afford more accurate risk models than traditional statistical models. These models can be used broadly to build parametric insurance for many types of risk and supplement traditional indemnity-based insurance by contributing to data analysis and risk estimations (Fig.~\ref{fig:overview}).

\section{Methods}

\begin{figure}[t]%
    \centering
    \includegraphics[width=0.8\textwidth]{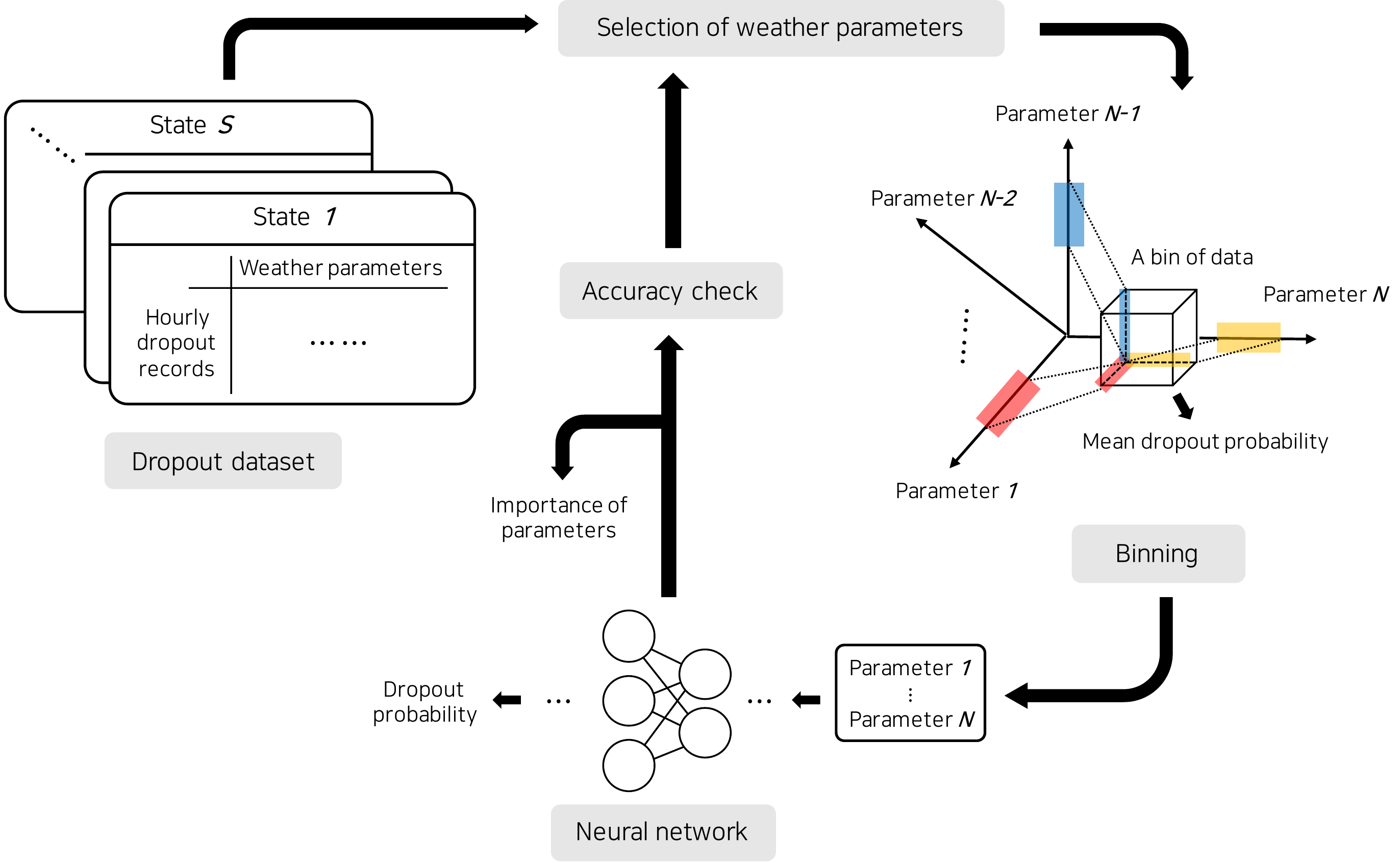}
    \label{fig:framework}
    \caption{A schematic showing the procedures on data pre-processing and analysis using neural networks.}
\end{figure}

Fig.~\ref{fig:framework} shows an overview on our method using the public dropout data and neural networks. 

\subsection{Data description}
For every state (State $1,\cdots,S$ in Fig.~\ref{fig:framework}) in US, the dataset provide information on the occurrence of dropout in residential network links. It contains the number of hours that the ThunderPing system checks if an address is responsive and that an address experienced dropout. Each check from the system is accompanied by a number of weather parameters, including temperature, wind speed, precipitation, snow depth, cloud ceiling, visibility, sea level, station pressure, etc. Among these parameters, we exclude ones with incomplete or invalid records.

\subsection{Data pre-processing} \label{sec:preprocess}

Dropouts are extreme events and are very rare in this dataset. Accordingly, they highly suffer from noises, which greatly hinders the training of neural networks. Hence, instead of estimating the occurrence of dropout from the raw data, we leverage some ideas suggested in \cite{padmanabhan2019residential}. First, the occurrence of dropouts is expressed in terms of the dropout probability, which is calculated as the number of hours to check responses divided by the number of hours that dropouts happen. Second, to suppress any undesirable effects from extreme cases, we use estimated mean of the dropout probability by considering bins of measurements from the ThunderPing system. Here, each bin corresponds to specific conditions of weather parameters $\theta_1, \cdots, \theta_n$. 

Specifically, for a parameter $\theta_i$, we suggest an interval $\left[ a_i - \Delta_i/2, a_i + \Delta_i/2 \right]$ where $a_i$ is a real number randomly chosen between the minimum $m_{i,\downarrow}$ and the maximum $m_{i,\uparrow}$ values of $\theta_i$ in a state. The interval width $\Delta_i$ is set to $(m_{i,\downarrow} - m_{i,\uparrow}) / 12$. Subsequently, a bin is defined as a collection of measurements whose parameters satisfy 
\begin{equation} \label{eq:bininterval}
    \theta_i \in 
    \left[ 
    a_i - \frac{\Delta_i}{2}, a_i + \frac{\Delta_i}{2} \right], 
    \,\, i \in \{1,\cdots,n \}.
\end{equation}
For each measurement $j$ in a bin, we calculate the dropout probability $p_{D,j}$. Denoting the number of measurements in a bin as $N$, we approximate the dropout probability for a bin $p_D$ by using the median of means of $p_{D,j}$, i.e.
\begin{equation} \label{eq:medmeans}
    p_D = \operatorname{med}
    \Biggl\{
    \frac{1}{q} \sum_{j=qk+1}^{q(k+1)} p_{D,j}
    \mathrel{\bigg\lvert}
    k \in \mathbb{N} \cup \{0\},\,\,
    q(k+1) \leq N
    \Biggr\},
\end{equation}
where $q$ is a natural number smaller than $N$. We only consider bins that satisfy $N\geq 4000$. $q$ is chosen such that the maximum of $k$ in Eq.~\ref{eq:medmeans} is larger than $20$. Consequently, we collect pairs $\left( \{a_1,\cdots,a_n\}, p_D \right)$ from multiple bins, as inputs to various models to estimate the dropout probability from weather parameters.

\subsection{Bayesian regression}
For a comparison to the estimation of the dropout probability using DSPPs, we perform the Bayesian regression as a conventional risk estimation scheme. To reflect the fact that the relationship between the weather parameters and the dropout probability can be nonlinear, we consider the following model,
\begin{equation}
    p_D = \mathcal{N}
    \left(
    w_0 
    + \sum_{i\in\mathbb{N}}^{0 < a_i+b_i+c_i \leq 2}
    w_i \theta_1^{a_i} \theta_2^{b_i} \theta_3^{c_i}
    \mathrel{\bigg\lvert}
    \sigma^2
    \right),
\end{equation}
where $a_i$, $b_i$, and $c_i$ are natural numbers, and $\mathcal{N}(\mu \lvert \sigma^2)$ is a normal distribution with the mean $\mu$ and the standard deviation $\sigma$. $\sigma$ is sampled from a half-normal distribution while each weight $w_i$ is sampled from a normal distribution. Here, we fix the number of weather parameters to 3, which is based on our initial selection of parameters (see, Section~\ref{sec:paramsel}). For the optimization of all model parameters, we perform the Monte Carlo simulation with $10^4$ samples and $6$ chains. To check the proper convergence, we confirm that the R-hat values for all model parameters are 1.

\subsection{Training of DSPP}
We fit a DSPP model to the pre-processed dataset. DSPP is a variant of the original DGP, as introduced earlier. In the training of DGP, the loss function is derived using the evidence lower bound due to the intractability of the model. In addition, the latent function values in each layer are randomly samples, which can make the inference difficult and thus degrade the predictive distribution. It is suggested that using DSPP can mitigate such issues \cite{jankowiak2020deep}.

Specifically, we approximate the variational distribution with 8 quadrature points and weights with quadrature rule referred to as QR3 in \cite{jankowiak2020deep}. Our network consists of 4 layers where the dimensions of their output vectors are 5, 3, 3, and 1. The covariance matrices in all layers are computed using the Mat\'ern covariance function and a scalar scaler. The mean vectors in the first 3 layers are modeled with the fully-connected layers. The last layer uses a constant mean. As in the training of the original DGP \cite{salimbeni2017doubly}, we use 300 inducing points for calculating the Kullback-Leibler divergence in all layers. Starting from an initial rate of 0.1, we train for 500 epochs for each state and decay the learning rate by 0.1 at the $100^{\text{th}}$, $250^{\text{th}}$, $350^{\text{th}}$, and $450^{\text{th}}$ epochs. We use a mini-batch of size 1000.

\subsection{Selection and importance of parameters} \label{sec:paramsel}

As discussed in Section~\ref{sec:preprocess}, each weather parameter suffers from noises and unexpected fluctuations. Hence, it is expected that using too many parameters (e.g. $n\gg1$) would result in a combination of different noises and undesirable effects in the training process. Furthermore, it can be difficult to find a bin with sufficiently many measurements, because the intersection of intervals (Eq.~\ref{eq:bininterval}) would be very small. Hence, at the initial stage of our method, we try to roughly pick a few among available parameters. This is done by calculating correlation values between the dropout probability and each parameter per measurement. Based on these correlation values, we select three parameters: temperature, wind speed, and precipitation. Detailed information regarding our selection is described in Appendix~\ref{app:choice}.

We expect that for parametric insurance, it would be useful to know the importance of weather parameters, i.e. determining a parameter that can most influence residential links. Observing the regional dependence of important features, we can know that location-specific insurance policies would be required in insurance products. In this study, we use a similar method to the feature ablation \cite{sheikholeslami2021autoablation,covert2020understanding,hooker2019please}. That is, making one weather parameter to be zero to remove any information flow regarding the parameter to a trained neural network and observing the accuracy change. We further train two additional DSPPs, denoted as DSPP$_1$ and DSPP$_2$, where the former is trained by removing the second most important parameter and the latter by removing the second and the third most important parameters. This is to further analyze the effects of weather parameters. Depending on the accuracy change, we may update our selection of parameters (Fig.~\ref{fig:overview}). In this study, we consider two accuracy measures, the mean absolute percentage error (MAPE) and the R-squared value. In particular, we use MAPE for the determination of the importance of parameters.

\section{Results}\label{sec:results}

\begin{figure}[t]%
    \centering
    \includegraphics[width=0.6\textwidth]{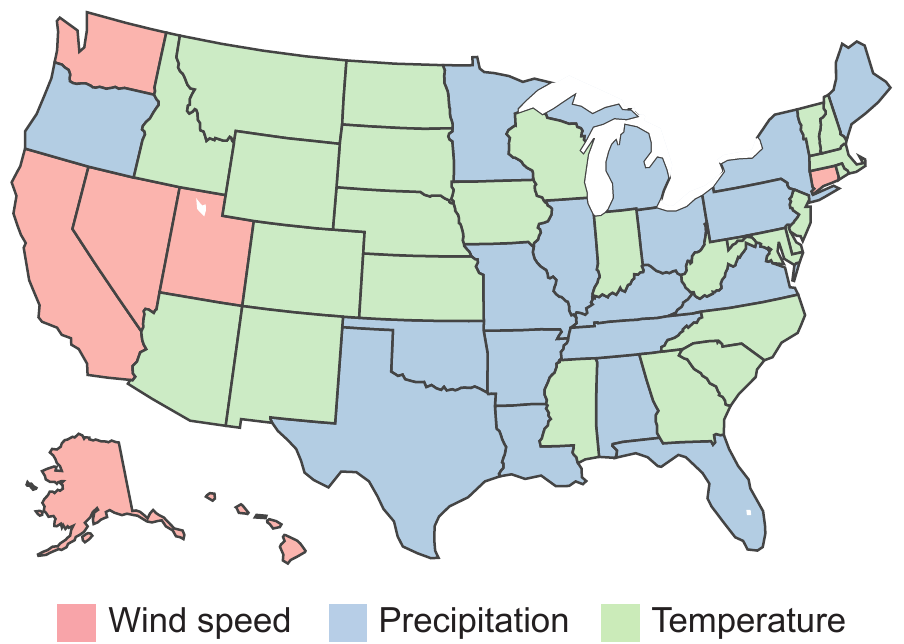}
    \caption{The most important weather parameter among wind speed, precipitation, and temperature for states in US.}
    \label{fig:parammap}
\end{figure}

\begin{figure}[t]%
    \centering
    \includegraphics[width=0.9\textwidth]{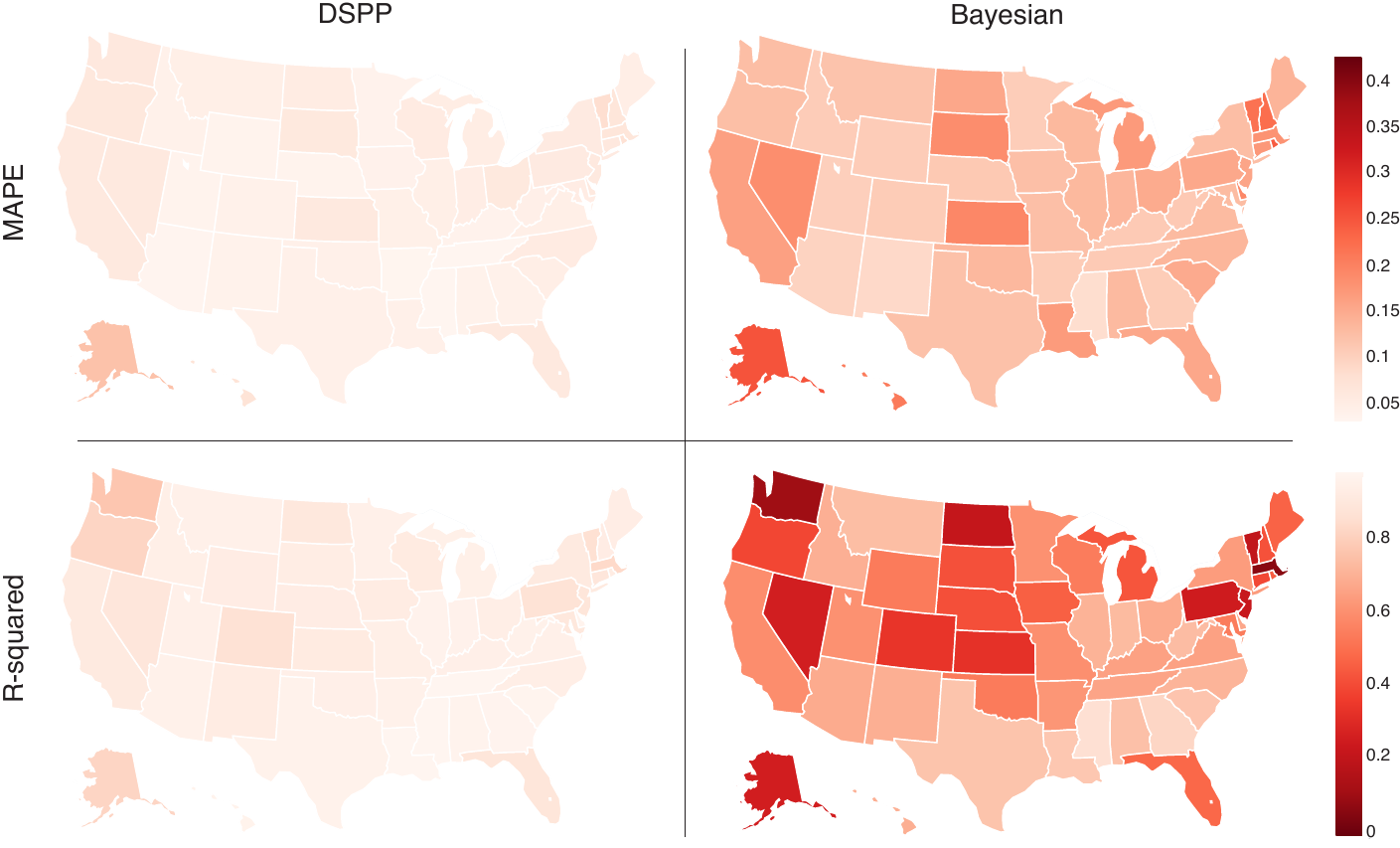}
    \caption{Accuracy of estimating the dropout probability by using (left column) the DSPP and (right column) the classical Bayesian regression. The range of the colorbars are same to that in Fig.~\ref{fig:abgcompmap} for better comparison.}
    \label{fig:compmap}
\end{figure}

\begin{figure}[t]%
    \centering
    \includegraphics[width=\textwidth]{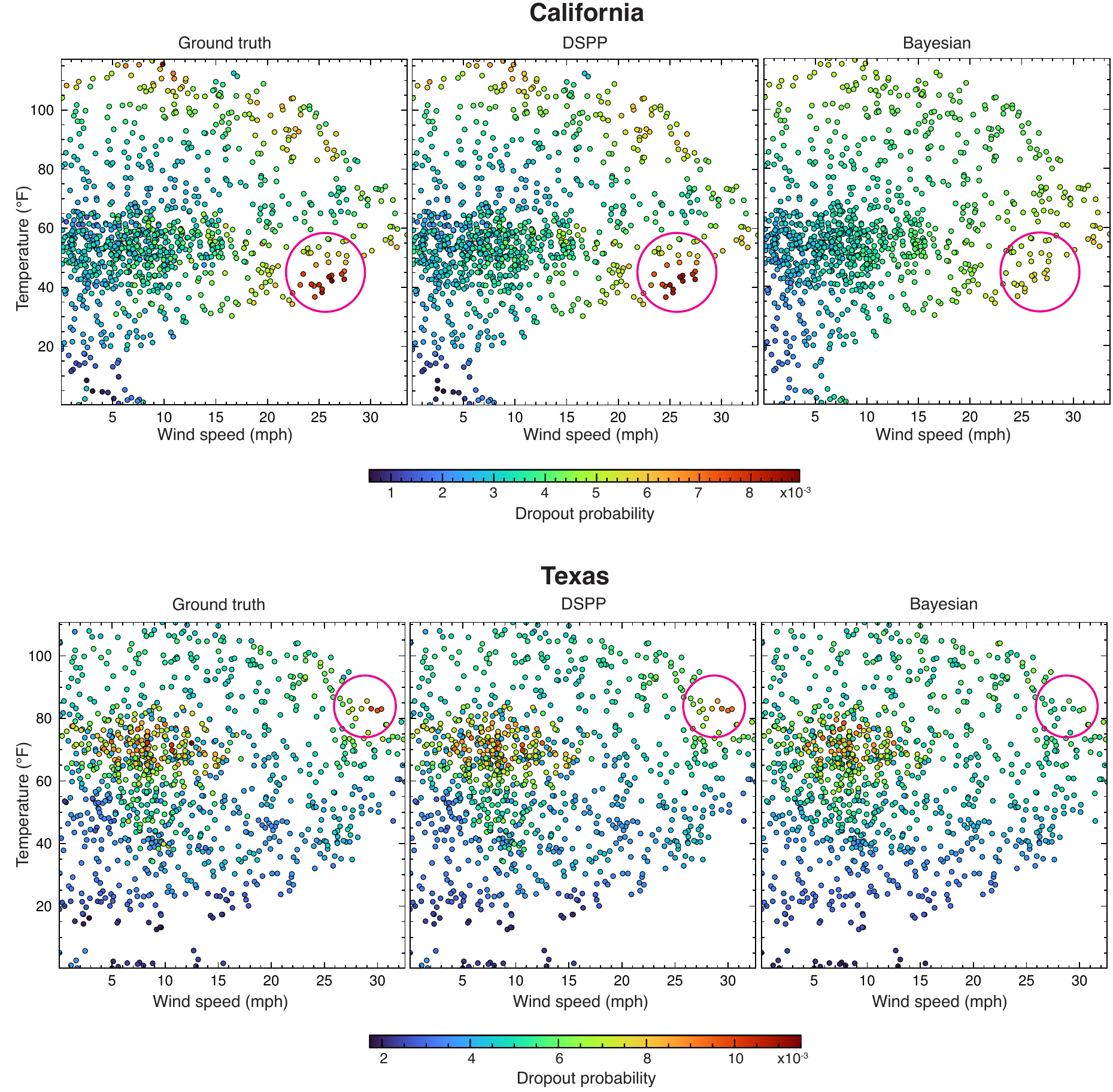}
    \caption{Estimation of the dropout probability in California and Texas using DSPP and Bayesian regression. For better visibility, we show two-dimensional (temperature and wind speed) view of the dropout probability.}
    \label{fig:twodimcomp}
\end{figure}

\begin{figure}[t]%
    \centering
    \includegraphics[width=\textwidth]{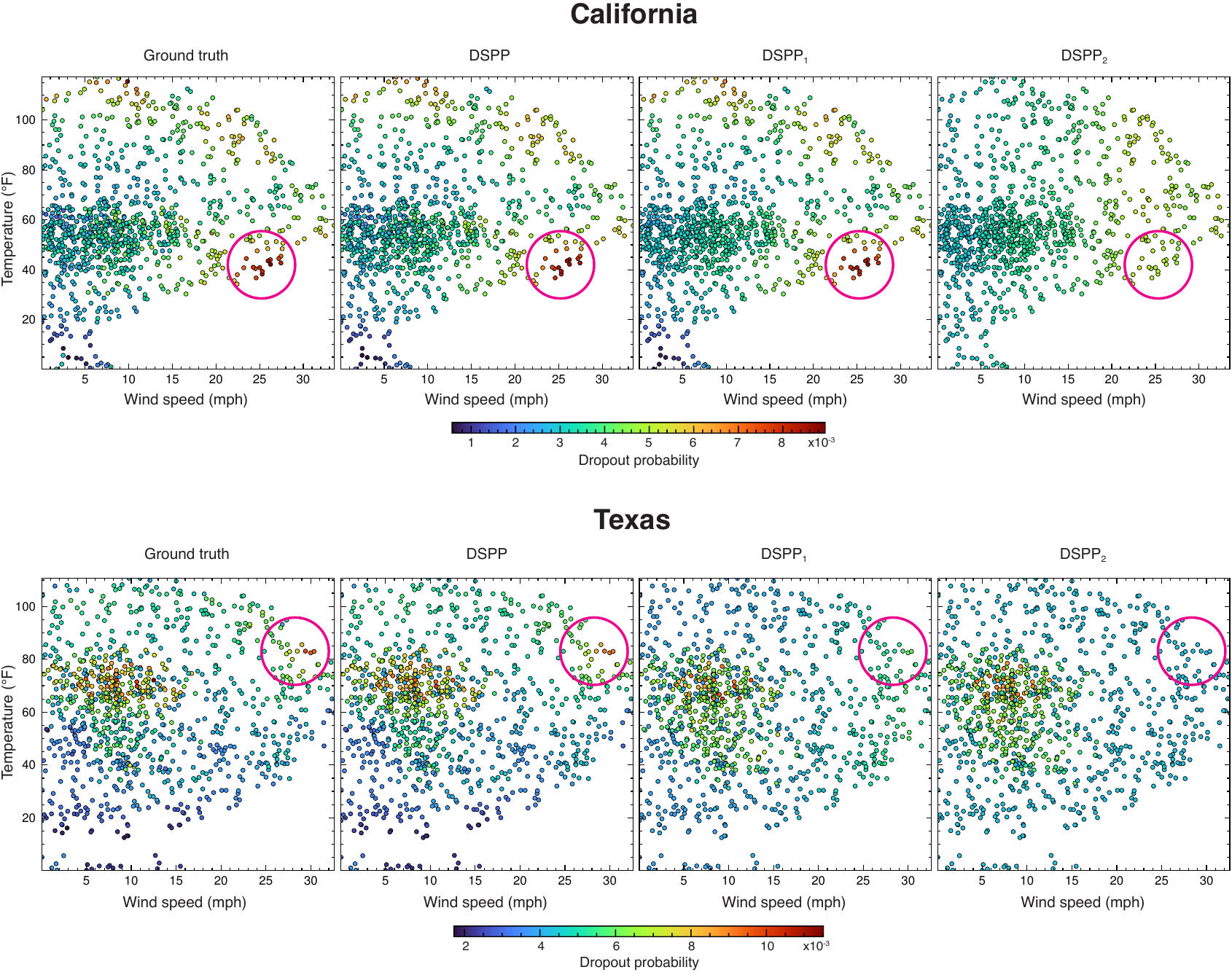}
    \caption{Estimation of the dropout probability in California and Texas using DSPP, DSPP$_1$, and DSPP$_2$. For better visibility, we show two-dimensional (temperature and wind speed) view of the dropout probability.}
    \label{fig:abgtwodimcomp}
\end{figure}

\begin{table}[!ht]
\begin{center}
\begin{minipage}{\textwidth}
\caption{MAPE on 1000 randomly sampled test data, using DSPP, DSPP$_1$, DSPP$_2$, and Bayesian regression. $a_1$, $a_2$, and $a_3$ represent the first, second, and third most important parameters, respectively. $W$, $P$, and $T$ refer to wind speed, precipitation, and temperature, respectively. Here we consider states from AK to MO. For the other states, see, Table~\ref{tab:latter}.}\label{tab:former}
\begin{tabular*}{\textwidth}{@{\extracolsep{\fill}}cccccccc@{\extracolsep{\fill}}}
\toprule%
State & DSPP & DSPP$_{1}$ & DSPP$_{2}$ & Bayesian & $a_1$ & $a_2$ & $a_3$ \\
\midrule
    AK & 0.122 & 0.177 & 0.273 & 0.254 & $W$ & $P$ & $T$ \\ 
    AL & 0.04 & 0.26 & 0.312 & 0.132 & $P$ & $T$ & $W$ \\ 
    AR & 0.033 & 0.177 & 0.208 & 0.108 & $P$ & $T$ & $W$ \\ 
    AZ & 0.034 & 0.074 & 0.202 & 0.098 & $T$ & $W$ & $P$ \\ 
    CA & 0.06 & 0.092 & 0.186 & 0.163 & $W$ & $P$ & $T$ \\ 
    CO & 0.042 & 0.106 & 0.11 & 0.108 & $T$ & $W$ & $P$ \\ 
    CT & 0.062 & 0.203 & 0.231 & 0.173 & $W$ & $T$ & $P$ \\ 
    DC & 0.043 & 0.106 & 0.217 & 0.09 & $T$ & $W$ & $P$ \\ 
    DE & 0.06 & 0.206 & 0.359 & 0.193 & $T$ & $W$ & $P$ \\ 
    FL & 0.06 & 0.137 & 0.164 & 0.155 & $P$ & $T$ & $W$ \\ 
    GA & 0.042 & 0.089 & 0.118 & 0.107 & $T$ & $P$ & $W$ \\ 
    HI & 0.073 & 0.291 & 0.348 & 0.208 & $W$ & $T$ & $P$ \\ 
    IA & 0.041 & 0.071 & 0.174 & 0.125 & $T$ & $P$ & $W$ \\ 
    ID & 0.042 & 0.12 & 0.223 & 0.108 & $T$ & $W$ & $P$ \\ 
    IL & 0.047 & 0.261 & 0.271 & 0.134 & $P$ & $T$ & $W$ \\ 
    IN & 0.05 & 0.088 & 0.294 & 0.136 & $T$ & $P$ & $W$ \\ 
    KS & 0.06 & 0.24 & 0.24 & 0.193 & $T$ & $W$ & $P$ \\ 
    KY & 0.04 & 0.182 & 0.184 & 0.111 & $P$ & $T$ & $W$ \\ 
    LA & 0.043 & 0.329 & 0.426 & 0.169 & $P$ & $T$ & $W$ \\ 
    MA & 0.069 & 0.12 & 0.188 & 0.179 & $T$ & $P$ & $W$ \\ 
    MD & 0.052 & 0.127 & 0.214 & 0.13 & $T$ & $W$ & $P$ \\ 
    ME & 0.049 & 0.131 & 0.176 & 0.14 & $P$ & $T$ & $W$ \\ 
    MI & 0.052 & 0.223 & 0.226 & 0.171 & $P$ & $T$ & $W$ \\ 
    MN & 0.039 & 0.172 & 0.184 & 0.107 & $P$ & $T$ & $W$ \\ 
    MO & 0.045 & 0.172 & 0.193 & 0.125 & $P$ & $T$ & $W$ \\
\botrule
\end{tabular*}
\end{minipage}
\end{center}
\end{table}

\begin{table}[!ht]
\begin{center}
\begin{minipage}{\textwidth}
\caption{MAPE on 1000 randomly sampled test data, using DSPP, DSPP$_1$, DSPP$_2$, and Bayesian regression. $a_1$, $a_2$, and $a_3$ represent the first, second, and third most important parameters, respectively. $W$, $P$, and $T$ refer to wind speed, precipitation, and temperature, respectively. Here we consider states from MS to WY. For the other states, see, Table~\ref{tab:former}.}\label{tab:latter}
\begin{tabular*}{\textwidth}{@{\extracolsep{\fill}}cccccccc@{\extracolsep{\fill}}}
\toprule%
State & DSPP & DSPP$_{1}$ & DSPP$_{2}$ & Bayesian & $a_1$ & $a_2$ & $a_3$ \\
\midrule
    MS & 0.032 & 0.05 & 0.297 & 0.084 & $T$ & $P$ & $W$ \\ 
    MT & 0.046 & 0.255 & 0.255 & 0.12 & $T$ & $W$ & $P$ \\ 
    NC & 0.055 & 0.101 & 0.274 & 0.137 & $T$ & $P$ & $W$ \\ 
    ND & 0.053 & 0.143 & 0.14 & 0.155 & $T$ & $W$ & $P$ \\ 
    NE & 0.038 & 0.116 & 0.113 & 0.113 & $T$ & $W$ & $P$ \\ 
    NH & 0.068 & 0.104 & 0.222 & 0.224 & $T$ & $P$ & $W$ \\ 
    NJ & 0.061 & 0.128 & 0.211 & 0.157 & $T$ & $W$ & $P$ \\ 
    NM & 0.04 & 0.084 & 0.179 & 0.09 & $T$ & $W$ & $P$ \\ 
    NV & 0.061 & 0.205 & 0.204 & 0.186 & $W$ & $T$ & $P$ \\ 
    NY & 0.057 & 0.176 & 0.211 & 0.127 & $P$ & $T$ & $W$ \\ 
    OH & 0.061 & 0.265 & 0.265 & 0.15 & $P$ & $T$ & $W$ \\ 
    OK & 0.043 & 0.174 & 0.202 & 0.134 & $P$ & $T$ & $W$ \\ 
    OR & 0.063 & 0.135 & 0.179 & 0.126 & $P$ & $T$ & $W$ \\ 
    PA & 0.058 & 0.162 & 0.173 & 0.153 & $P$ & $T$ & $W$ \\ 
    RI & 0.077 & 0.255 & 0.36 & 0.244 & $T$ & $W$ & $P$ \\ 
    SC & 0.048 & 0.095 & 0.139 & 0.151 & $T$ & $P$ & $W$ \\ 
    SD & 0.061 & 0.19 & 0.178 & 0.186 & $T$ & $W$ & $P$ \\ 
    TN & 0.033 & 0.246 & 0.244 & 0.111 & $P$ & $T$ & $W$ \\ 
    TX & 0.044 & 0.181 & 0.192 & 0.122 & $P$ & $T$ & $W$ \\ 
    UT & 0.037 & 0.161 & 0.163 & 0.104 & $W$ & $T$ & $P$ \\ 
    VA & 0.047 & 0.198 & 0.215 & 0.131 & $P$ & $T$ & $W$ \\ 
    VT & 0.083 & 0.194 & 0.267 & 0.219 & $T$ & $W$ & $P$ \\ 
    WA & 0.062 & 0.103 & 0.13 & 0.127 & $W$ & $P$ & $T$ \\ 
    WI & 0.055 & 0.089 & 0.21 & 0.133 & $T$ & $P$ & $W$ \\ 
    WV & 0.046 & 0.095 & 0.244 & 0.112 & $T$ & $P$ & $W$ \\ 
    WY & 0.036 & 0.114 & 0.175 & 0.108 & $T$ & $W$ & $P$ \\ 
\botrule
\end{tabular*}
\end{minipage}
\end{center}
\end{table}

Based on our trained DSPP, we estimate the importance of parameters. Subsequently, we find that each state has a unique weather parameter that is most effective on the rate of dropout (Fig.~\ref{fig:parammap}). In other words, in Fig.~\ref{fig:parammap}, we can see that states differed in terms of whether wind speed, precipitation, or temperature is the strongest predictor of internet connectivity dropout.

In terms of modeling accuracy, DSPPs demonstrate significantly better performance than the Bayesian regression, as shown in Fig.~\ref{fig:compmap}. Specifically, we calculate the MAPE and the R-squared value by using 1000 randomly selected test examples for each state. Depending on the state, the change in the accuracy differs (see also Table~\ref{tab:former}~and~\ref{tab:latter}). This may be because extra non-linearity considered in DSPPs compared to the Bayesian regression model has different impact in various states. In overall, the distribution of the dropout probability depends on the state, implying that the regional dependency should be considered.

For more detailed analysis, Fig.~\ref{fig:twodimcomp} shows two-dimensional (temperature and wind speed) views of the dropout probability estimated from DSPP and Bayesian regression in California and Texas. In particular, the temperature and wind speed with the highest probability of dropout are marked with circles. Qualitatively, estimation from DSPP exhibits strong agreement to the ground truth, with high dropout probability for the weather conditions in the circled regions. However, the classical Bayesian regression does not well approximate the high probability in those regions, implying less performance in catching some extreme events. In overall, our DSPP has more accurate risk modeling than the classical Bayesian regression. In Appendix~\ref{app:regfig}, Fig.~\ref{fig:threedimcomp} shows three-dimensional ($z$-axis also representing the dropout probability) views with uncertainty estimations, which also exhibits clear distinction between performances for DSPP and the Bayesian regression.

Fig.~\ref{fig:abgtwodimcomp} shows a comparison between estimations from DSPP, DSPP$_1$, and DSPP$_2$ in California and Texas. Here, we also mark regions exhibiting the highest dropout probability with circles. In the example of California, the removal of the second most important parameter does not have a significant effect on the performance. This would be because information in the second most important parameter is suitably retained in the other two parameters, while direct relationship between parameters may require additional analysis. On the other hand, in Texas, the removal of the second most important parameter results in a significant change in the estimation quality. Such degradation in the estimation quality implies that, in Texas, the effect of the third important parameter is relatively small compared to the case in California. These observations can suggest the importance of AI-based data analysis that can assess diverse data with geographical differences. In Appendix~\ref{app:regfig}, Figs.~\ref{fig:abgthreedimcompca}~and\ref{fig:abgthreedimcomptx} provide three-dimensional views with uncertainty estimations in California and Texas. In these figures, we can also see different behaviors on omitting parameters, depending on states.

Quantitative results in the estimation of the dropout probability are summarized in Table~\ref{tab:former}~and~\ref{tab:latter} where we show MAPE from 1000 randomly sampled test data regarding the classial Bayesian regression, DSPP, DSPP$_1$, and DSPP$_2$. In overall, estimations from DSPP exhibit MAPE of approximately 5\%, which is the best result among the other models. As in Fig.~\ref{fig:parammap}, we can see the regional dependency of the importance of parameters. As in the previous discussion, neglecting the second and the third most important parameters has distinct effects on the estimation accuracy. For more visibility, we show Fig.~\ref{fig:abgcompmap} that contains maps of estimation quality for DSPP$_1$, and DSPP$_2$. Taken together, these results suggest that our DSPP can predict risk with greater accuracy than traditional models, which can be applied to the data analysis portion of parametric insurance for more accurate coverage than traditional insurance can achieve.

\section{Discussion} \label{sec:discussion}

In addition to the better estimation quality, our model can provide uncertainty on its estimation, which is relevant to not only being more robust to overfitting but the insurance design. Being highly uncertain implies that, for example, there exists insufficient amount of training data on which estimations of a model are based. On the other hand, it can be also possible that the relationship between risks and parameters is complex. 

There is no rule-of-thumb to quantify the quality of uncertainty estimation \cite{li2021deep}, but we can suggest a few qualitative ways. For example, models with high uncertainty almost everywhere do not give useful information. In addition, we can expect that models should exhibit high uncertainty on weather conditions with a small number of measurements. In this perspective, Figs.~\ref{fig:threedimcomp},~\ref{fig:abgthreedimcompca},~and~\ref{fig:abgthreedimcomptx} suggest that our DSPP model is also better in terms of uncertainty estimation compared to other methods. The Bayesian regression model and DSPPs without some parameters show high uncertainty on various measurement points. In contrast, uncertainty values from our DSPP model tends to be proportional to the density of measurement points, which follows our qualitative expectation. We expect that such information on uncertainty can be crucial in insurance business. We can know that it is more probable for real-world phenomena to deviate from expectations of models if uncertainty is high, and we would require more careful and conservative policy on such expectations or additional data to tailor current models.




Parametric insurance models consist of not only data analysis and risk estimation procedures, but considerations on the trigger and payouts (Fig.~\ref{fig:overview}). There can be multiple ways to apply our study to build parametric insurance products. Roughly speaking, the expected amount of risk is highly related with the dropout probability. Considering the maximum limit for reserve and number of policyholders, we may devise thresholds on the dropout probability for which we can consider triggering payouts. Subsequently, dropout probabilities, estimated from our model, can be clustered into several regions. Regions where the minimum dropout probability is larger than some thresholds give crucial information on triggers, i.e. we can consider releasing payouts if all weather parameters are contained in such regions. 

While insurance is one of the primary ways for societies to protect themselves from financial risk, insurance companies may not be able to afford to insure climate-change-related risk. Many studies focus on the impacts of climate on insurance companies’ viability. For example, \cite{hawker2007climate} highlights that insurance companies are vulnerable to increasing financial losses with climate change, which implies that better models and data analysis procedures are required. In this regard, we expect that this study can contribute to insurance designs in the future.

\section{Conclusion}\label{sec:conclusion}

Parametric insurance has attracted attention due to its advantages over traditional insurance products in some aspects, such as faster and easier claim procedures. One of the main problems in parametric insurance, however, is the problem on estimating the basis risk. In this paper, we study the potential use of Bayesian neural networks to reduce the basis risk. Specifically, we use data regarding the dropouts in residential links in US, which are influenced by weather conditions. We show that, by using DSPPs, we can estimate the dropout probability with significantly less amount of error compared to the traditional Bayesian regression models. In addition, we can see clear regional dependence of the importance of weather parameters, which implies that we should consider geographical differences in designing parametric insurance. 

Consequently, we expect that our methods can be applied not only to building data-driven parametric insurance models for internet connectivity dropout in US but to building parametric insurance models for a multitude of risks that conventional insurance cannot adequately cover.

\backmatter

\section*{Declarations}
The authors disclose no conflicts of interest. 

\begin{appendices}

\section{Initial choice of parameters via correlation} \label{app:choice}

\begin{figure}[t]%
    \centering
    \includegraphics[width=\textwidth]{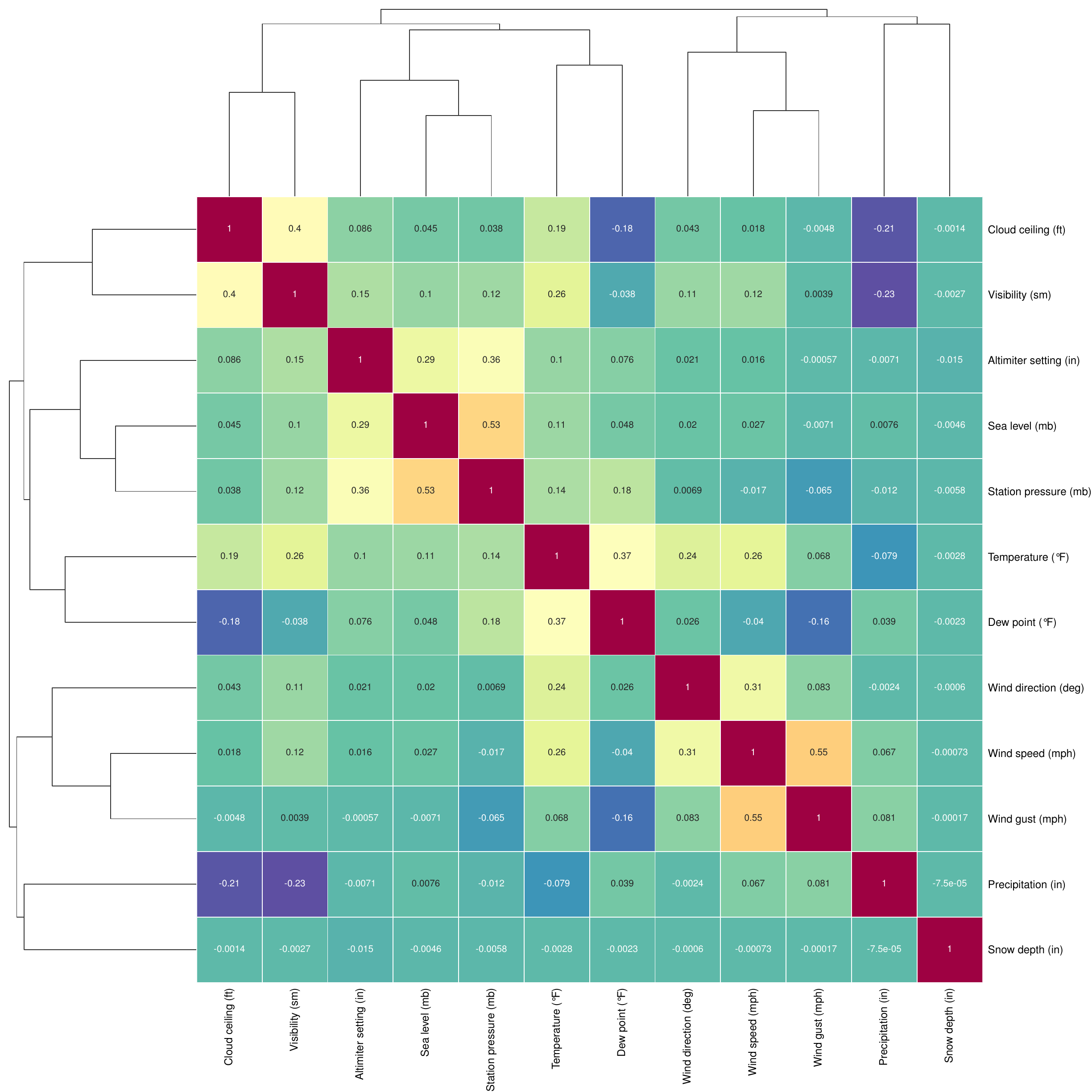}
    \caption{Correlation matrix and hierarchical clustering of weather parameters via the UPGMA algorithm in California.}
    \label{fig:clusterca}
\end{figure}

\begin{figure}[t]%
    \centering
    \includegraphics[width=\textwidth]{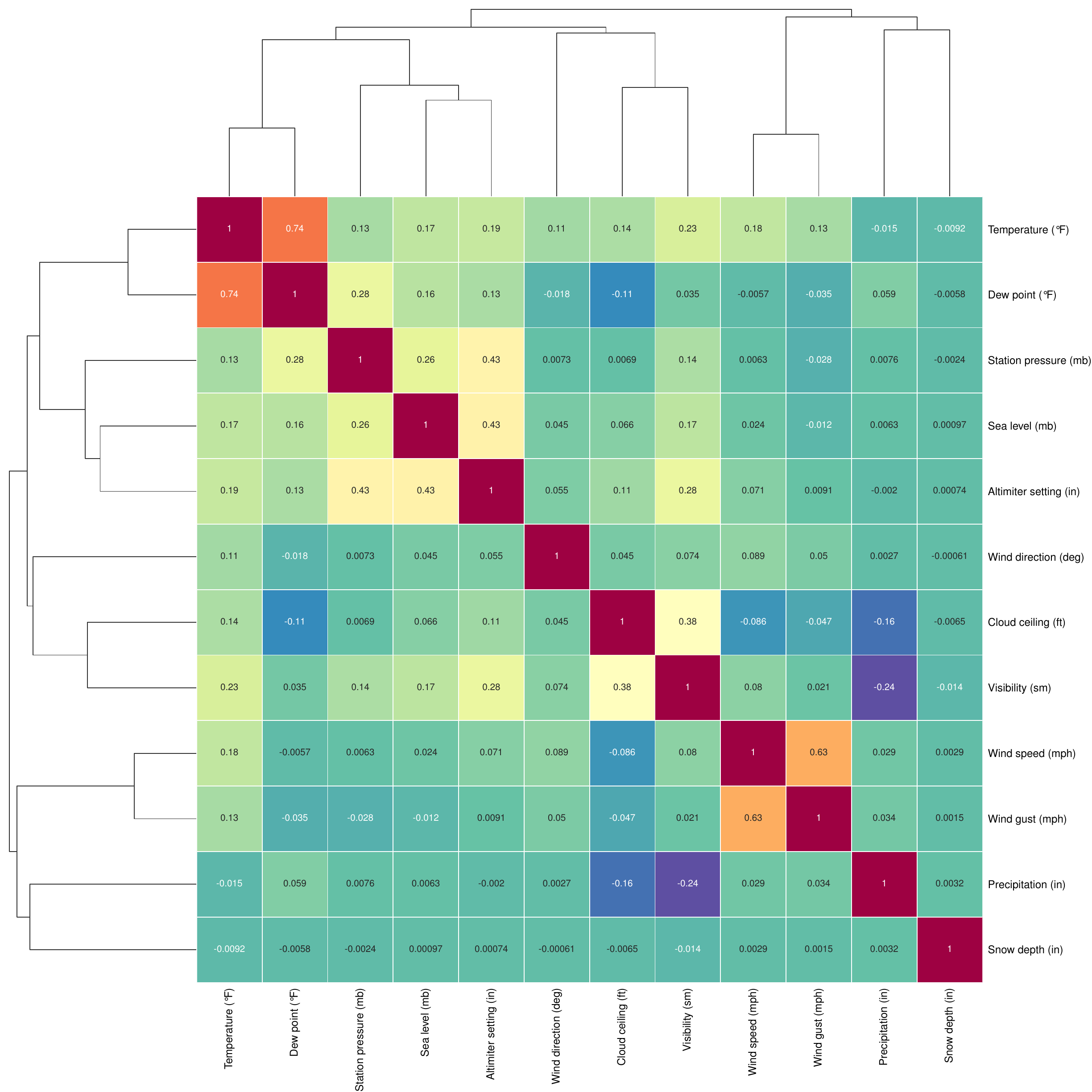}
    \caption{Correlation matrix and hierarchical clustering of weather parameters via the UPGMA algorithm in Texas.}
    \label{fig:clustertx}
\end{figure}

As illustrated in Sec.~\ref{sec:paramsel}, using all of available weather parameters would be problematic, because each parameter is subjective to its own noise. Subsequently, it is important to select a few weather parameters that can well describe the dropout. However, it would be difficult to know the effects of weather parameters on the dropout before we actually collect measurements into multiple bins and train models. Therefore, to obtain an initial guess on a set of appropriate weather parameters, we construct correlation matrices and try to select a few representative parameters.

As an example, Figs.~\ref{fig:clusterca}~and~\ref{fig:clustertx} show correlation matrices between the dropout probability and various weather parameters. They also contain hierarchical clustering results computed by using the unweighted pair group method with arithmetic mean (UPGMA) algorithm \citep{jain1988algorithms}. To guarantee sufficient amount of information regarding the dropout, it would be beneficial to select parameters that are orthogonal to each other. In both states, we have a few clusters: (cloud ceiling, visibility), (temperature, dew point), (station pressure, sea level, altimeter setting), (wind speed, wind gust), precipitation, wind direction, and snow depth. Considering general applicability among these parameters, e.g. wind direction may be arbitrary and snow depth may be inapplicable in certain states, we select three common weather parameters from the clustering results, precipitation, wind speed, and temperature, as our initial selection.

\section{Additional figures for estimation results in California and Texas} \label{app:regfig}

\begin{figure}[t]%
    \centering
    \includegraphics[width=\textwidth]{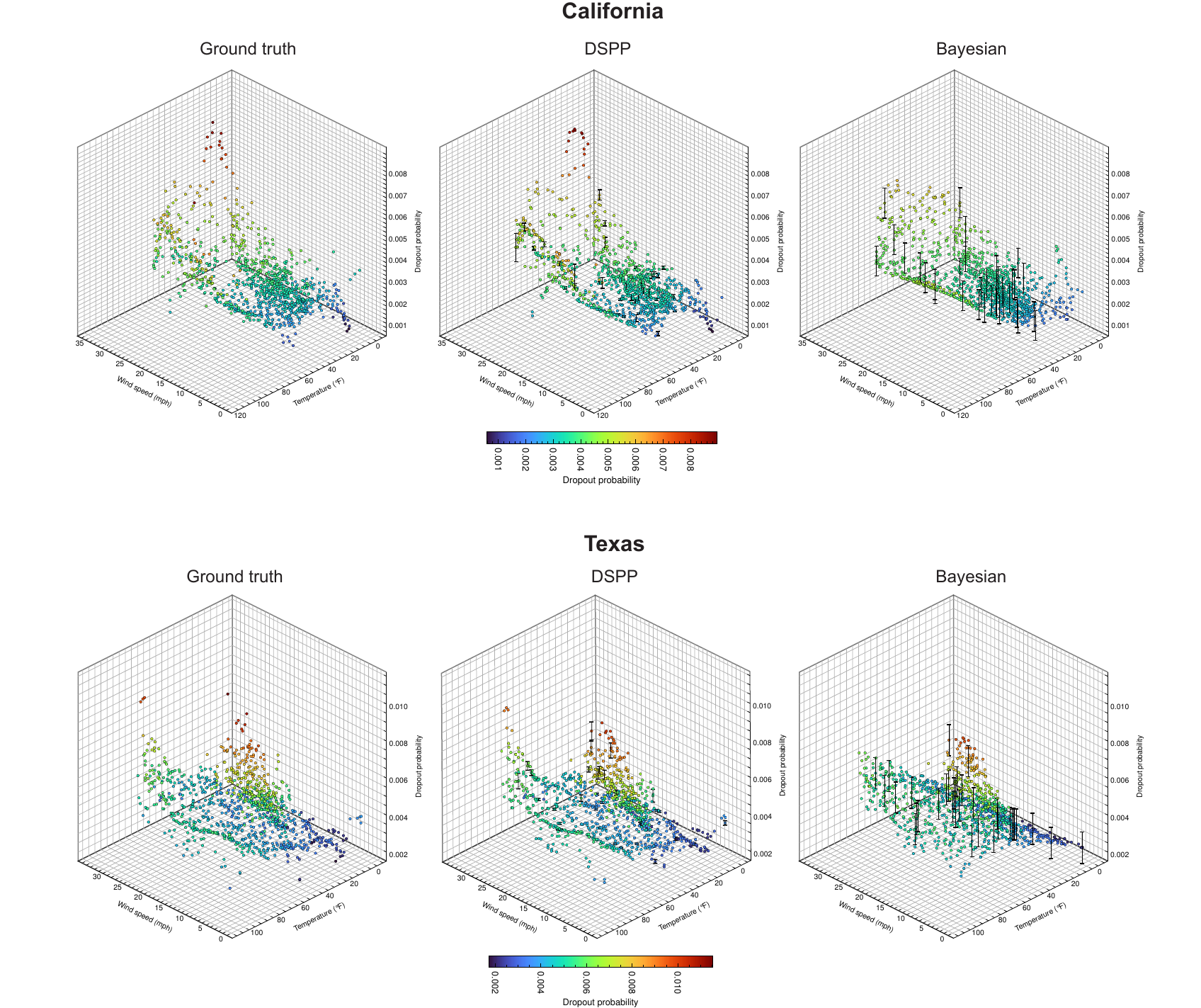}
    \caption{Three-dimensional comparison between estimations from DSPP and classical Bayesian regression in California and Texas. Note that both the color and $z$-axis values represent the dropout probability. Black errorbars are drawn at randomly selected regions, which represent uncertainty on estimations.}
    \label{fig:threedimcomp}
\end{figure}

\begin{figure}[t]%
    \centering
    \includegraphics[width=0.9\textwidth]{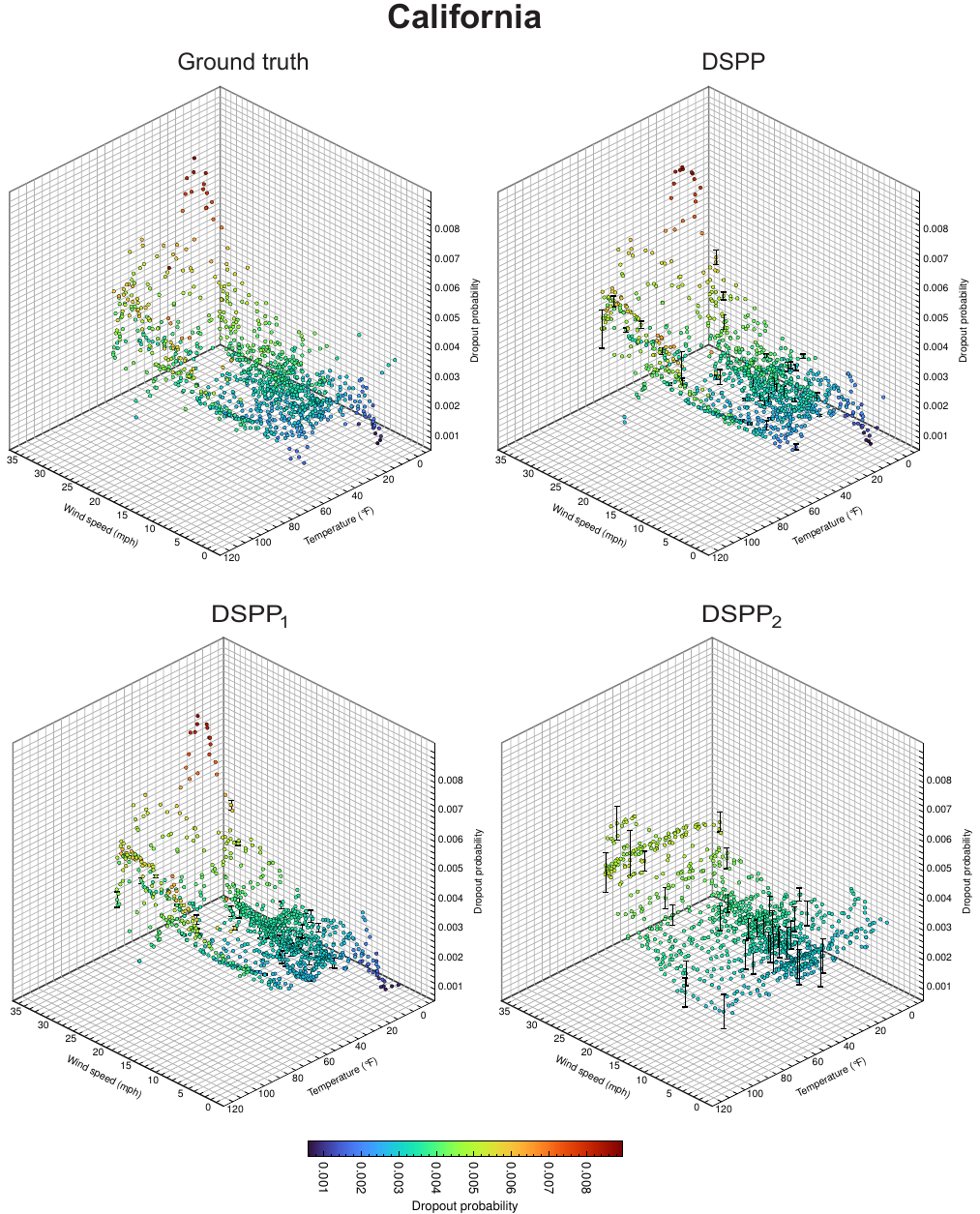}
    \caption{Three-dimensional comparison between estimations from DSPP, DSPP$_1$, and DSPP$_2$ in California. Note that both the color and $z$-axis values represent the dropout probability. Black errorbars are drawn at randomly selected regions, which represent uncertainty on estimations.}
    \label{fig:abgthreedimcompca}
\end{figure}

\begin{figure}[t]%
    \centering
    \includegraphics[width=0.9\textwidth]{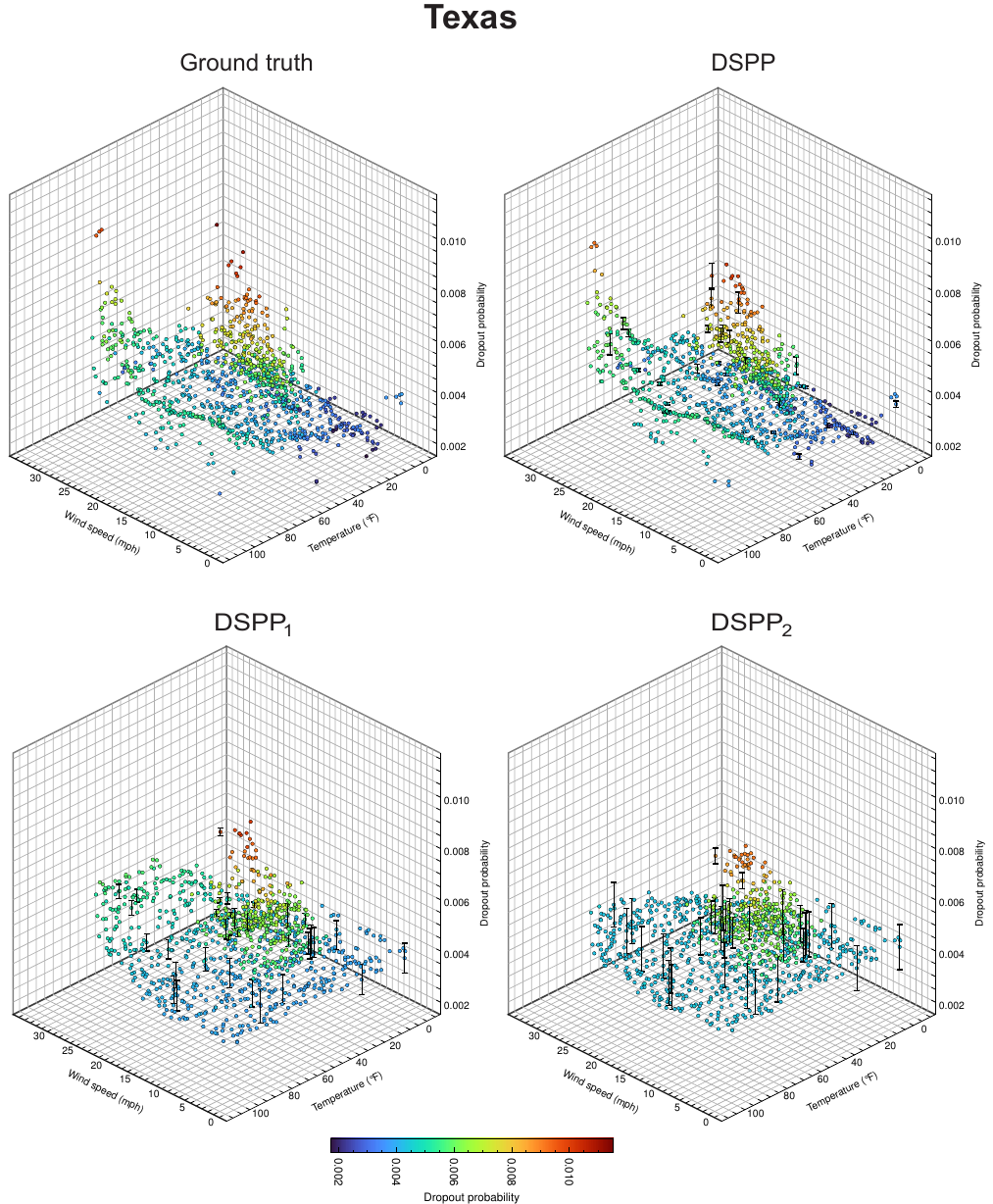}
    \caption{Three-dimensional comparison between estimations from DSPP, DSPP$_1$, and DSPP$_2$ in Texas. Note that both the color and $z$-axis values represent the dropout probability. Black errorbars are drawn at randomly selected regions, which represent uncertainty on estimations.}
    \label{fig:abgthreedimcomptx}
\end{figure}

In this section, we show three-dimensional views of estimations from Bayesian regression, DSPP, DSPP$_1$, and DSPP$_2$ models. Here, in these views, note that both the color and $z$-axis values represent the dropout probability; for example, points with high $z$-value will have red color and low $z$-value points will have blue color. This is to emphasize differences between estimations more. In these views, we also draw black errorbars at randomly selected regions, which correspond to uncertainty on estimations.

As discussed in Sections~\ref{sec:results}~and~\ref{sec:discussion}, it can be clearly seen that DSPP can best approximate the ground truth probability. In some cases, such as classical Bayesian regression and DSPP$_2$, we can see high uncertainty values almost everywhere, implying that uncertainty estimation is not successful.

\section{Maps of estimation accuracy using DSPP$_1$ and DSPP$_2$} \label{app:dspp12}

\begin{figure}[t]%
    \centering
    \includegraphics[width=\textwidth]{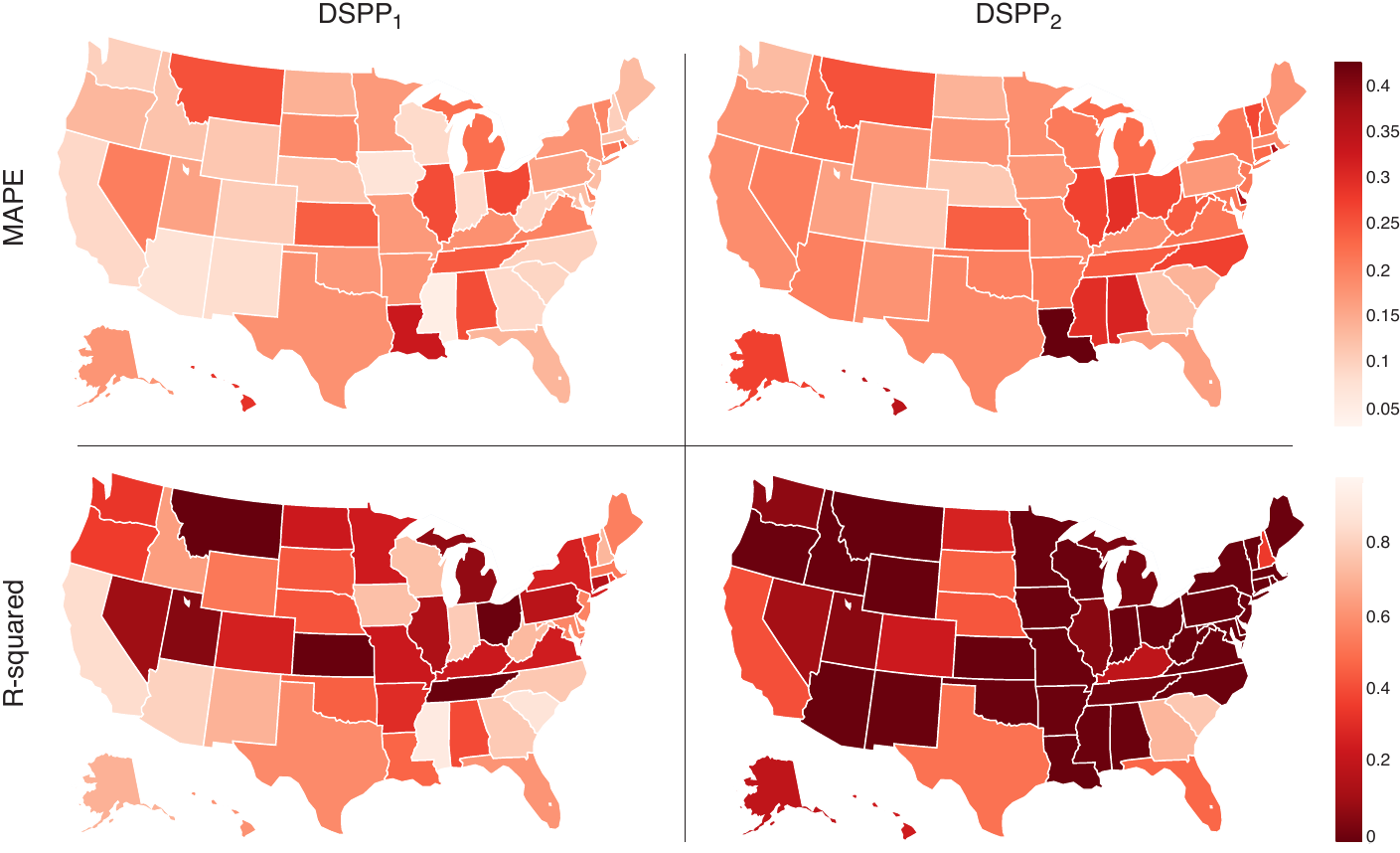}
    \caption{Accuracy of estimating the dropout probability by using (left column) DSPP$_1$ and (right column) DSPP$_2$.}
    \label{fig:abgcompmap}
\end{figure}

In addition to quantitative results summarized in Tables~\ref{tab:former}~and~\ref{tab:latter}, we show maps of estimation errors from DSPPs without a part of parameters: DSPP$_1$ and DSPP$_2$. 

\end{appendices}

\bibliography{refs}


\begin{thebibliography}{39}
\ifx \bisbn   \undefined \def \bisbn  #1{ISBN #1}\fi
\ifx \binits  \undefined \def \binits#1{#1}\fi
\ifx \bauthor  \undefined \def \bauthor#1{#1}\fi
\ifx \batitle  \undefined \def \batitle#1{#1}\fi
\ifx \bjtitle  \undefined \def \bjtitle#1{#1}\fi
\ifx \bvolume  \undefined \def \bvolume#1{\textbf{#1}}\fi
\ifx \byear  \undefined \def \byear#1{#1}\fi
\ifx \bissue  \undefined \def \bissue#1{#1}\fi
\ifx \bfpage  \undefined \def \bfpage#1{#1}\fi
\ifx \blpage  \undefined \def \blpage #1{#1}\fi
\ifx \burl  \undefined \def \burl#1{\textsf{#1}}\fi
\ifx \doiurl  \undefined \def \doiurl#1{\url{https://doi.org/#1}}\fi
\ifx \betal  \undefined \def \betal{\textit{et al.}}\fi
\ifx \binstitute  \undefined \def \binstitute#1{#1}\fi
\ifx \binstitutionaled  \undefined \def \binstitutionaled#1{#1}\fi
\ifx \bctitle  \undefined \def \bctitle#1{#1}\fi
\ifx \beditor  \undefined \def \beditor#1{#1}\fi
\ifx \bpublisher  \undefined \def \bpublisher#1{#1}\fi
\ifx \bbtitle  \undefined \def \bbtitle#1{#1}\fi
\ifx \bedition  \undefined \def \bedition#1{#1}\fi
\ifx \bseriesno  \undefined \def \bseriesno#1{#1}\fi
\ifx \blocation  \undefined \def \blocation#1{#1}\fi
\ifx \bsertitle  \undefined \def \bsertitle#1{#1}\fi
\ifx \bsnm \undefined \def \bsnm#1{#1}\fi
\ifx \bsuffix \undefined \def \bsuffix#1{#1}\fi
\ifx \bparticle \undefined \def \bparticle#1{#1}\fi
\ifx \barticle \undefined \def \barticle#1{#1}\fi
\bibcommenthead
\ifx \bconfdate \undefined \def \bconfdate #1{#1}\fi
\ifx \botherref \undefined \def \botherref #1{#1}\fi
\ifx \url \undefined \def \url#1{\textsf{#1}}\fi
\ifx \bchapter \undefined \def \bchapter#1{#1}\fi
\ifx \bbook \undefined \def \bbook#1{#1}\fi
\ifx \bcomment \undefined \def \bcomment#1{#1}\fi
\ifx \oauthor \undefined \def \oauthor#1{#1}\fi
\ifx \citeauthoryear \undefined \def \citeauthoryear#1{#1}\fi
\ifx \endbibitem  \undefined \def \endbibitem {}\fi
\ifx \bconflocation  \undefined \def \bconflocation#1{#1}\fi
\ifx \arxivurl  \undefined \def \arxivurl#1{\textsf{#1}}\fi
\csname PreBibitemsHook\endcsname

\bibitem{chester2020keeping}
\begin{barticle}
\bauthor{\bsnm{Chester}, \binits{M.V.}},
\bauthor{\bsnm{Underwood}, \binits{B.S.}},
\bauthor{\bsnm{Samaras}, \binits{C.}}:
\batitle{Keeping infrastructure reliable under climate uncertainty}.
\bjtitle{Nature Climate Change}
\bvolume{10}(\bissue{6}),
\bfpage{488}--\blpage{490}
(\byear{2020})
\end{barticle}
\endbibitem

\bibitem{neumann2015climate}
\begin{barticle}
\bauthor{\bsnm{Neumann}, \binits{J.E.}},
\bauthor{\bsnm{Price}, \binits{J.}},
\bauthor{\bsnm{Chinowsky}, \binits{P.}},
\bauthor{\bsnm{Wright}, \binits{L.}},
\bauthor{\bsnm{Ludwig}, \binits{L.}},
\bauthor{\bsnm{Streeter}, \binits{R.}},
\bauthor{\bsnm{Jones}, \binits{R.}},
\bauthor{\bsnm{Smith}, \binits{J.B.}},
\bauthor{\bsnm{Perkins}, \binits{W.}},
\bauthor{\bsnm{Jantarasami}, \binits{L.}}, \betal:
\batitle{Climate change risks to {US} infrastructure: impacts on roads,
  bridges, coastal development, and urban drainage}.
\bjtitle{Climatic Change}
\bvolume{131}(\bissue{1}),
\bfpage{97}--\blpage{109}
(\byear{2015})
\end{barticle}
\endbibitem

\bibitem{yates2014stormy}
\begin{barticle}
\bauthor{\bsnm{Yates}, \binits{D.}},
\bauthor{\bsnm{Luna}, \binits{B.Q.}},
\bauthor{\bsnm{Rasmussen}, \binits{R.}},
\bauthor{\bsnm{Bratcher}, \binits{D.}},
\bauthor{\bsnm{Garre}, \binits{L.}},
\bauthor{\bsnm{Chen}, \binits{F.}},
\bauthor{\bsnm{Tewari}, \binits{M.}},
\bauthor{\bsnm{Friis-Hansen}, \binits{P.}}:
\batitle{Stormy weather: {A}ssessing climate change hazards to electric power
  infrastructure: {A} sandy case study}.
\bjtitle{IEEE Power and Energy Magazine}
\bvolume{12}(\bissue{5}),
\bfpage{66}--\blpage{75}
(\byear{2014})
\end{barticle}
\endbibitem

\bibitem{falco2014crop}
\begin{barticle}
\bauthor{\bsnm{Falco}, \binits{S.D.}},
\bauthor{\bsnm{Adinolfi}, \binits{F.}},
\bauthor{\bsnm{Bozzola}, \binits{M.}},
\bauthor{\bsnm{Capitanio}, \binits{F.}}:
\batitle{Crop insurance as a strategy for adapting to climate change}.
\bjtitle{Journal of Agricultural Economics}
\bvolume{65}(\bissue{2}),
\bfpage{485}--\blpage{504}
(\byear{2014})
\end{barticle}
\endbibitem

\bibitem{ji2022impact}
\begin{barticle}
\bauthor{\bsnm{Ji}, \binits{T.}},
\bauthor{\bsnm{Yao}, \binits{Y.}},
\bauthor{\bsnm{Dou}, \binits{Y.}},
\bauthor{\bsnm{Deng}, \binits{S.}},
\bauthor{\bsnm{Yu}, \binits{S.}},
\bauthor{\bsnm{Zhu}, \binits{Y.}},
\bauthor{\bsnm{Liao}, \binits{H.}}:
\batitle{The impact of climate change on urban transportation resilience to
  compound extreme events}.
\bjtitle{Sustainability}
\bvolume{14}(\bissue{7}),
\bfpage{3880}
(\byear{2022})
\end{barticle}
\endbibitem

\bibitem{jorgensen2020natural}
\begin{barticle}
\bauthor{\bsnm{J{\o}rgensen}, \binits{S.L.}},
\bauthor{\bsnm{Termansen}, \binits{M.}},
\bauthor{\bsnm{Pascual}, \binits{U.}}:
\batitle{Natural insurance as condition for market insurance: {C}limate change
  adaptation in agriculture}.
\bjtitle{Ecological Economics}
\bvolume{169},
\bfpage{106489}
(\byear{2020})
\end{barticle}
\endbibitem

\bibitem{national2008potential}
\begin{bbook}
\bauthor{\bsnm{{Committee on Climate Change and {U.S.} Transportation, National
  Research Council}}}:
\bbtitle{Potential Impacts of Climate Change on {U.S.} Transportation}.
\bpublisher{National Research Council},
\blocation{U.S.}
(\byear{2008})
\end{bbook}
\endbibitem

\bibitem{wang2012impact}
\begin{barticle}
\bauthor{\bsnm{Wang}, \binits{X.}},
\bauthor{\bsnm{Stewart}, \binits{M.G.}},
\bauthor{\bsnm{Nguyen}, \binits{M.}}:
\batitle{Impact of climate change on corrosion and damage to concrete
  infrastructure in {A}ustralia}.
\bjtitle{Climatic Change}
\bvolume{110}(\bissue{3}),
\bfpage{941}--\blpage{957}
(\byear{2012})
\end{barticle}
\endbibitem

\bibitem{wilbanks2013climate}
\begin{bbook}
\bauthor{\bsnm{Wilbanks}, \binits{T.}},
\bauthor{\bsnm{Fernandez}, \binits{S.}}:
\bbtitle{Climate Change and Infrastructure, Urban Systems, and
  Vulnerabilities}.
\bpublisher{Island Press},
\blocation{Washington, DC}
(\byear{2014})
\end{bbook}
\endbibitem

\bibitem{authority2015impact}
\begin{bbook}
\bauthor{\bsnm{{Prudential Regulation Authority}}}:
\bbtitle{The Impact of Climate Change on the UK Insurance Sector}.
\bpublisher{Bank of England},
\blocation{London}
(\byear{2015})
\end{bbook}
\endbibitem

\bibitem{botzen2010climate}
\begin{barticle}
\bauthor{\bsnm{Botzen}, \binits{W.J.W.}},
\bauthor{\bparticle{van~den} \bsnm{Bergh}, \binits{J.C.J.M.}},
\bauthor{\bsnm{Bouwer}, \binits{L.M.}}:
\batitle{Climate change and increased risk for the insurance sector: a global
  perspective and an assessment for the {N}etherlands}.
\bjtitle{Natural Hazards}
\bvolume{52}(\bissue{3}),
\bfpage{577}--\blpage{598}
(\byear{2010})
\end{barticle}
\endbibitem

\bibitem{mills2007synergisms}
\begin{barticle}
\bauthor{\bsnm{Mills}, \binits{E.}}:
\batitle{Synergisms between climate change mitigation and adaptation: an
  insurance perspective}.
\bjtitle{Mitigation and Adaptation Strategies for Global Change}
\bvolume{12}(\bissue{5}),
\bfpage{809}--\blpage{842}
(\byear{2007})
\end{barticle}
\endbibitem

\bibitem{collier2021climate}
\begin{barticle}
\bauthor{\bsnm{Collier}, \binits{S.J.}},
\bauthor{\bsnm{Elliott}, \binits{R.}},
\bauthor{\bsnm{Lehtonen}, \binits{T.-K.}}:
\batitle{Climate change and insurance}.
\bjtitle{Economy and Society}
\bvolume{50}(\bissue{2}),
\bfpage{158}--\blpage{172}
(\byear{2021})
\end{barticle}
\endbibitem

\bibitem{dlugolecki2008climate}
\begin{barticle}
\bauthor{\bsnm{Dlugolecki}, \binits{A.}}:
\batitle{Climate change and the insurance sector}.
\bjtitle{The Geneva Papers on Risk and Insurance-Issues and Practice}
\bvolume{33}(\bissue{1}),
\bfpage{71}--\blpage{90}
(\byear{2008})
\end{barticle}
\endbibitem

\bibitem{hecht2007climate}
\begin{barticle}
\bauthor{\bsnm{Hecht}, \binits{S.B.}}:
\batitle{Climate change and the transformation of risk: {I}nsurance matters}.
\bjtitle{UCLA Law Review}
\bvolume{55},
\bfpage{1559}
(\byear{2007})
\end{barticle}
\endbibitem

\bibitem{mills2009global}
\begin{barticle}
\bauthor{\bsnm{Mills}, \binits{E.}}:
\batitle{A global review of insurance industry responses to climate change}.
\bjtitle{The Geneva Papers on Risk and Insurance-Issues and Practice}
\bvolume{34}(\bissue{3}),
\bfpage{323}--\blpage{359}
(\byear{2009})
\end{barticle}
\endbibitem

\bibitem{kraehnert2021insurance}
\begin{barticle}
\bauthor{\bsnm{Kraehnert}, \binits{K.}},
\bauthor{\bsnm{Osberghaus}, \binits{D.}},
\bauthor{\bsnm{Hott}, \binits{C.}},
\bauthor{\bsnm{Habtemariam}, \binits{L.T.}},
\bauthor{\bsnm{W{\"a}tzold}, \binits{F.}},
\bauthor{\bsnm{Hecker}, \binits{L.P.}},
\bauthor{\bsnm{Fluhrer}, \binits{S.}}:
\batitle{Insurance against extreme weather events: {A}n overview}.
\bjtitle{Review of Economics}
\bvolume{72}(\bissue{2}),
\bfpage{71}--\blpage{95}
(\byear{2021})
\end{barticle}
\endbibitem

\bibitem{pan2022assessing}
\begin{barticle}
\bauthor{\bsnm{Pan}, \binits{Q.}},
\bauthor{\bsnm{Porth}, \binits{L.}},
\bauthor{\bsnm{Li}, \binits{H.}}:
\batitle{Assessing the effectiveness of the actuaries climate index for
  estimating the impact of extreme weather on crop yield and insurance
  applications}.
\bjtitle{Sustainability}
\bvolume{14}(\bissue{11}),
\bfpage{6916}
(\byear{2022})
\end{barticle}
\endbibitem

\bibitem{lawrence2020leveraging}
\begin{barticle}
\bauthor{\bsnm{Lawrence}, \binits{J.-M.}},
\bauthor{\bsnm{Hossain}, \binits{N.U.I.}},
\bauthor{\bsnm{Jaradat}, \binits{R.}},
\bauthor{\bsnm{Hamilton}, \binits{M.}}:
\batitle{Leveraging a {B}ayesian network approach to model and analyze supplier
  vulnerability to severe weather risk: {A} case study of the {US}
  pharmaceutical supply chain following {H}urricane {M}aria}.
\bjtitle{International Journal of Disaster Risk Reduction}
\bvolume{49},
\bfpage{101607}
(\byear{2020})
\end{barticle}
\endbibitem

\bibitem{ma2019flash}
\begin{barticle}
\bauthor{\bsnm{Ma}, \binits{M.}},
\bauthor{\bsnm{Liu}, \binits{C.}},
\bauthor{\bsnm{Zhao}, \binits{G.}},
\bauthor{\bsnm{Xie}, \binits{H.}},
\bauthor{\bsnm{Jia}, \binits{P.}},
\bauthor{\bsnm{Wang}, \binits{D.}},
\bauthor{\bsnm{Wang}, \binits{H.}},
\bauthor{\bsnm{Hong}, \binits{Y.}}:
\batitle{Flash flood risk analysis based on machine learning techniques in the
  {Y}unnan {P}rovince, {C}hina}.
\bjtitle{Remote Sensing}
\bvolume{11}(\bissue{2}),
\bfpage{170}
(\byear{2019})
\end{barticle}
\endbibitem

\bibitem{cesarini2021potential}
\begin{barticle}
\bauthor{\bsnm{Cesarini}, \binits{L.}},
\bauthor{\bsnm{Figueiredo}, \binits{R.}},
\bauthor{\bsnm{Monteleone}, \binits{B.}},
\bauthor{\bsnm{Martina}, \binits{M.L.}}:
\batitle{The potential of machine learning for weather index insurance}.
\bjtitle{Natural Hazards and Earth System Sciences}
\bvolume{21}(\bissue{8}),
\bfpage{2379}--\blpage{2405}
(\byear{2021})
\end{barticle}
\endbibitem

\bibitem{porrini2014insurance}
\begin{barticle}
\bauthor{\bsnm{Porrini}, \binits{D.}},
\bauthor{\bsnm{Schwarze}, \binits{R.}}:
\batitle{Insurance models and european climate change policies: an assessment}.
\bjtitle{European Journal of Law and Economics}
\bvolume{38},
\bfpage{7}--\blpage{28}
(\byear{2014})
\end{barticle}
\endbibitem

\bibitem{mahul2010government}
\begin{bbook}
\bauthor{\bsnm{Mahul}, \binits{O.}},
\bauthor{\bsnm{Stutley}, \binits{C.J.}}:
\bbtitle{Government Support to Agricultural Insurance: Challenges and Options
  for Developing Countries}.
\bpublisher{World Bank},
\blocation{Washington, DC}
(\byear{2010})
\end{bbook}
\endbibitem

\bibitem{vroege2019index}
\begin{barticle}
\bauthor{\bsnm{Vroege}, \binits{W.}},
\bauthor{\bsnm{Dalhaus}, \binits{T.}},
\bauthor{\bsnm{Finger}, \binits{R.}}:
\batitle{Index insurances for grasslands -– {A} review for {E}urope and
  {N}orth-{A}merica}.
\bjtitle{Agricultural Systems}
\bvolume{168},
\bfpage{101}--\blpage{111}
(\byear{2019})
\end{barticle}
\endbibitem

\bibitem{miranda2012index}
\begin{barticle}
\bauthor{\bsnm{Miranda}, \binits{M.J.}},
\bauthor{\bsnm{Farrin}, \binits{K.}}:
\batitle{Index insurance for developing countries}.
\bjtitle{Applied Economic Perspectives and Policy}
\bvolume{34}(\bissue{3}),
\bfpage{391}--\blpage{427}
(\byear{2012})
\end{barticle}
\endbibitem

\bibitem{jankowiak2020deep}
\begin{bchapter}
\bauthor{\bsnm{Jankowiak}, \binits{M.}},
\bauthor{\bsnm{Pleiss}, \binits{G.}},
\bauthor{\bsnm{Gardner}, \binits{J.}}:
\bctitle{Deep sigma point processes}.
In: \bbtitle{Conference on Uncertainty in Artificial Intelligence},
pp. \bfpage{789}--\blpage{798}
(\byear{2020}).
\bcomment{Proceedings of Machine Learning Research}
\end{bchapter}
\endbibitem

\bibitem{damianou2013deep}
\begin{bchapter}
\bauthor{\bsnm{Damianou}, \binits{A.}},
\bauthor{\bsnm{Lawrence}, \binits{N.D.}}:
\bctitle{Deep {G}aussian processes}.
In: \bbtitle{Artificial Intelligence and Statistics},
pp. \bfpage{207}--\blpage{215}
(\byear{2013}).
\bcomment{Proceedings of Machine Learning Research}
\end{bchapter}
\endbibitem

\bibitem{cutajar2018deep}
\begin{botherref}
\oauthor{\bsnm{Cutajar}, \binits{K.}},
\oauthor{\bsnm{Pullin}, \binits{M.}},
\oauthor{\bsnm{Damianou}, \binits{A.}},
\oauthor{\bsnm{Lawrence}, \binits{N.}},
\oauthor{\bsnm{Gonzalez}, \binits{J.}}:
Deep {G}aussian processes for multi-fidelity modeling.
Advances in Neural Information Processing Systems, Third Workshop on Bayesian
  Deep Learning
(2018)
\end{botherref}
\endbibitem

\bibitem{foong2020expressiveness}
\begin{barticle}
\bauthor{\bsnm{Foong}, \binits{A.}},
\bauthor{\bsnm{Burt}, \binits{D.}},
\bauthor{\bsnm{Li}, \binits{Y.}},
\bauthor{\bsnm{Turner}, \binits{R.}}:
\batitle{On the expressiveness of approximate inference in {B}ayesian neural
  networks}.
\bjtitle{Advances in Neural Information Processing Systems}
\bvolume{33},
\bfpage{15897}--\blpage{15908}
(\byear{2020})
\end{barticle}
\endbibitem

\bibitem{damianou2015deep}
\begin{botherref}
\oauthor{\bsnm{Damianou}, \binits{A.}}:
Deep gaussian processes and variational propagation of uncertainty.
PhD thesis,
University of Sheffield
(2015)
\end{botherref}
\endbibitem

\bibitem{schulman2011pingin}
\begin{bchapter}
\bauthor{\bsnm{Schulman}, \binits{A.}},
\bauthor{\bsnm{Spring}, \binits{N.}}:
\bctitle{Pingin'in the rain}.
In: \bbtitle{Proceedings of the 2011 ACM SIGCOMM Conference on Internet
  Measurement Conference},
pp. \bfpage{19}--\blpage{28}
(\byear{2011})
\end{bchapter}
\endbibitem

\bibitem{padmanabhan2019residential}
\begin{bchapter}
\bauthor{\bsnm{Padmanabhan}, \binits{R.}},
\bauthor{\bsnm{Schulman}, \binits{A.}},
\bauthor{\bsnm{Levin}, \binits{D.}},
\bauthor{\bsnm{Spring}, \binits{N.}}:
\bctitle{Residential links under the weather}.
In: \bbtitle{Proceedings of the ACM Special Interest Group on Data
  Communication},
pp. \bfpage{145}--\blpage{158}
(\byear{2019})
\end{bchapter}
\endbibitem

\bibitem{salimbeni2017doubly}
\begin{botherref}
\oauthor{\bsnm{Salimbeni}, \binits{H.}},
\oauthor{\bsnm{Deisenroth}, \binits{M.}}:
Doubly stochastic variational inference for deep {G}aussian processes.
Advances in Neural Information Processing Systems
\textbf{30}
(2017)
\end{botherref}
\endbibitem

\bibitem{sheikholeslami2021autoablation}
\begin{bchapter}
\bauthor{\bsnm{Sheikholeslami}, \binits{S.}},
\bauthor{\bsnm{Meister}, \binits{M.}},
\bauthor{\bsnm{Wang}, \binits{T.}},
\bauthor{\bsnm{Payberah}, \binits{A.H.}},
\bauthor{\bsnm{Vlassov}, \binits{V.}},
\bauthor{\bsnm{Dowling}, \binits{J.}}:
\bctitle{Autoablation: {A}utomated parallel ablation studies for deep
  learning}.
In: \bbtitle{Proceedings of the 1st Workshop on Machine Learning and Systems},
pp. \bfpage{55}--\blpage{61}
(\byear{2021})
\end{bchapter}
\endbibitem

\bibitem{covert2020understanding}
\begin{barticle}
\bauthor{\bsnm{Covert}, \binits{I.}},
\bauthor{\bsnm{Lundberg}, \binits{S.M.}},
\bauthor{\bsnm{Lee}, \binits{S.-I.}}:
\batitle{Understanding global feature contributions with additive importance
  measures}.
\bjtitle{Advances in Neural Information Processing Systems}
\bvolume{33},
\bfpage{17212}--\blpage{17223}
(\byear{2020})
\end{barticle}
\endbibitem

\bibitem{hooker2019please}
\begin{botherref}
\oauthor{\bsnm{Hooker}, \binits{G.}},
\oauthor{\bsnm{Mentch}, \binits{L.}}:
Please stop permuting features: An explanation and alternatives.
arXiv preprint arXiv:1905.03151
(2019)
\end{botherref}
\endbibitem

\bibitem{li2021deep}
\begin{barticle}
\bauthor{\bsnm{Li}, \binits{Y.}},
\bauthor{\bsnm{Rao}, \binits{S.}},
\bauthor{\bsnm{Hassaine}, \binits{A.}},
\bauthor{\bsnm{Ramakrishnan}, \binits{R.}},
\bauthor{\bsnm{Canoy}, \binits{D.}},
\bauthor{\bsnm{Salimi-Khorshidi}, \binits{G.}},
\bauthor{\bsnm{Mamouei}, \binits{M.}},
\bauthor{\bsnm{Lukasiewicz}, \binits{T.}},
\bauthor{\bsnm{Rahimi}, \binits{K.}}:
\batitle{Deep {B}ayesian {G}aussian processes for uncertainty estimation in
  electronic health records}.
\bjtitle{Scientific Reports}
\bvolume{11}(\bissue{1}),
\bfpage{1}--\blpage{13}
(\byear{2021})
\end{barticle}
\endbibitem

\bibitem{hawker2007climate}
\begin{barticle}
\bauthor{\bsnm{Hawker}, \binits{M.}}:
\batitle{Climate change and the global insurance industry}.
\bjtitle{The Geneva Papers on Risk and Insurance-Issues and Practice}
\bvolume{32}(\bissue{1}),
\bfpage{22}--\blpage{28}
(\byear{2007})
\end{barticle}
\endbibitem

\bibitem{jain1988algorithms}
\begin{bbook}
\bauthor{\bsnm{Jain}, \binits{A.K.}},
\bauthor{\bsnm{Dubes}, \binits{R.C.}}:
\bbtitle{Algorithms for Clustering Data}.
\bpublisher{Prentice Hall},
\blocation{US}
(\byear{1988})
\end{bbook}
\endbibitem

\end{thebibliography}


\end{document}